\documentclass[smallcondensed]{svjour3}     
\def\makeheadbox{\relax} 

\smartqed  
\usepackage{graphicx}
\usepackage{hyperref}

\usepackage{mathptmx}      

\AtBeginDocument{%
  \paperwidth=\dimexpr
    1in + \oddsidemargin
    + \textwidth
    + 1in + \oddsidemargin
  \relax
  \paperheight=\dimexpr
    1in + \topmargin
    + \headheight + \headsep
    + \textheight
    + 1in + \topmargin
  \relax
  \usepackage[pass]{geometry}\relax
}

\usepackage{amsmath}

\usepackage{amsthm}
\usepackage{mathtools} 
\usepackage{dsfont} 

\DeclareMathSymbol{\mlq}{\mathord}{operators}{``}
\DeclareMathSymbol{\mrq}{\mathord}{operators}{`'}

\theoremstyle{plain}
\newtheorem{thm}{Theorem}
\newtheorem{ass}{Assumption}

\newtheorem{innercustomass}{Assumption}
\newenvironment{customass}[1]
  {\renewcommand\theinnercustomass{#1}\innercustomass}
  {\endinnercustomass}

\newcommand{\NSD}{\textsc{No-Superdeterminism}}
\newcommand{\LOC}{\textsc{Locality}}
\newcommand{\AOE}{\textsc{Absoluteness of Observed Events}}
\newcommand{\LF}{\textsc{Local Friendliness}}
\newcommand{\LA}{\textsc{Local Action}}
\newcommand{\TR}{\textsc{true}}
\newcommand{\FA}{\textsc{false}}

\newcommand{\ket}[1]{|#1\rangle}
\newcommand{\idt}{\mathds{1}}

\begin{document}

\title{The view from a Wigner bubble
}


\author{Eric G. Cavalcanti}


\institute{E. G. Cavalcanti \at
              Centre for Quantum Dynamics, Griffith University, Gold Coast, QLD 4222, Australia \\
              \email{e.cavalcanti@griffith.edu.au}           
}

\date{Feb 26, 2021}

\maketitle

\begin{abstract}
In a recent no-go theorem~[Bong \emph{et al.}, Nature Physics (2020)], we proved that the predictions of unitary quantum mechanics for an extended Wigner's friend scenario are incompatible with any theory satisfying three metaphysical assumptions, the conjunction of which we call ``Local Friendliness'': \AOE, \LOC~and \NSD. In this paper (based on an invited talk for the QBism jubilee at the 2019 V\"axj\"o conference\footnote{This article should be considered part of the \href{https://link.springer.com/article/10.1007/s10701-020-00401-0}{December 2020 special issue of Foundations of Physics: Quantum Information: Impact to Foundations}.}) I discuss the implications of this theorem for QBism, as seen from the point of view of experimental metaphysics. I argue that the key distinction between QBism and realist interpretations of quantum mechanics is best understood in terms of their adherence to different theories of truth: the pragmatist versus the correspondence theories. I argue that a productive pathway to resolve the measurement problem within a pragmatist view involves taking seriously the perspective of quantum betting agents, even those in what I call a ``Wigner bubble''. The notion of reality afforded by QBism, I propose, will correspond to the invariant elements of any theory that has pragmatic value to all rational agents---that is, the elements that are invariant upon changes of agent perspectives. The classical notion of `event' is not among those invariants, even when those events are observed by some agent. Neither are quantum states. Nevertheless, I argue that far from solipsism, a personalist view of quantum states is an expression of its precise opposite: \emph{Copernicanism}.
\end{abstract}


\section{Introduction}
\label{sec:Intro}

Firstly I should say that I don’t consider myself to be a QBist. If anything, I prefer to think of myself as an {\it experimental metaphysicist}\footnote{I am aware that philosophers who work on metaphysics are called ``metaphysicians''. I use ``metaphysicist'' to refer to physicists who work on (experimental) metaphysics.}~\cite{Cavalcanti2007t}, following the term introduced by Abner Shimony~\cite{Shimony1989} to refer to the field that has Bell's 1964 theorem~\cite{Bell1964} as its pioneering result. That is, I am interested in studying the landscape of metaphysically possible theories, and finding ways to rule out classes of such theories via theorems and experiments. The objects of study of experimental metaphysics are physical theories (thus the prefix ``meta'')\footnote{In the same sense as ``metamathematics''---the field that has mathematical theories as its object of study, and G\"odel's incompleteness theorem as its most prominent result.} and their relationship to the world of observation (thus the adjective ``experimental''). 

In relating theory to observation, one must extract the predictions of the theory from the perspective of users of the theory. This is called the ``operational theory''. QBism plays a special role among interpretations of quantum mechanics for being the quantum operational theory \emph{par excellence}. In general, two or more theories in the same operational equivalence class can have distinct ontologies---that is, they have different commitments about what elements of the theory, if any, are taken to be real and objective. Although most physicists have strong opinions about one or another ontological theory, experimental metaphysics aims to remain agnostic about philosophical commitments that are not implied by theorems or by experiment.

There's a story, I don’t know if it's apocryphal, but it's a nice one, according to which Michelangelo, when asked about how he sculpted David, said: ``I just removed anything that was not David''. I like to think of the metaphysical landscape as the raw block of marble---with different points in the block corresponding to different physical theories---and of experimental metaphysics as a chisel to carve the marble, eliminating corners that do not describe the world of our experience. It may turn out that we are unable to reduce the block to a single point, corresponding to the one true ``theory of everything''\footnote{In fact, this is most likely impossible, a fact that philosophers refer to as the ``underdetermination of theory by evidence''. For example, any set of publicly shared observations can always be described within a deterministic single-world theory, simply by adding a sufficient number of hidden variables and allowing the laws of physics to have a sufficiently complex mathematical form. There would no doubt be reasons to reject ad-hoc theories of this kind via some philosophical criteria, but they cannot be ruled out as \emph{impossible}.}. But we may hope that after we carve out all the bits that experiment allows us to, what remains forms a beautiful whole\footnote{Note that I am not advocating beauty as a criterion for truth, a view that has been recently criticised as unproductive for physics~\cite{Hossenfelder2018}. Every theory in the landscape is probably ``ugly'' by some physicists' aesthetic judgement. I'll return to the question of truth later in this paper.}. To appreciate and understand this landscape, a theorist must be comfortable with ``wearing the hat'' of different interpretations from time to time. 
\begin{figure}
    \centering
    \includegraphics[width=0.5\linewidth]{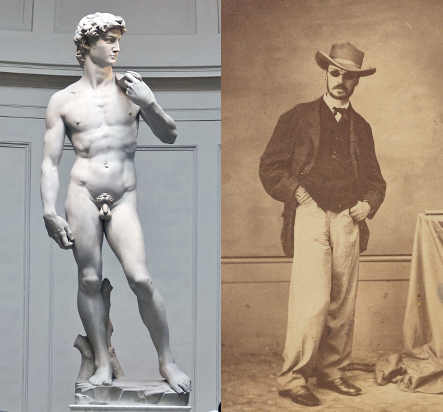}
\caption{Left: a representation of the landscape of physical theories. Right: William James, one of the leading proponents of American pragmatism, and a philosophical inspiration for QBism,  during a visit to Brazil in 1865, and the hat I wish I was wearing while delivering the talk this paper is based on.}
    \label{fig:concept_fig}
\end{figure}

In this paper I write with my QBist hat on. I want to contribute an argument against a charge commonly levied at QBism, that its insistence in interpreting quantum states as degrees of belief of individual agents amounts to {\it solipsism}. That is, to the idea that the only thing that exists is my own mind (or at any rate, the mind of whoever entertains solipsism). I will argue that, on the contrary, the personalist view of quantum states can be justified from a premise that is the {\it precise opposite} of solipsism, namely {\it Copernicanism}---the idea that no individual agent is at the center of the universe. I also argue that taking this idea seriously requires a radical revision of the classical notion of ``event'', and by extension, of space-time.

The story that I want to elaborate on here goes back to an argument I proposed to Chris Fuchs in Vienna in 2007. I recounted the argument in an email to Chris on February 2008, which was published on the arxiv in 2014~\cite{Fuchs2014}, as part of the second instalment of his ``samizdat'' series. For context, I reproduce it below:

\begin{quotation}
\textbf{Cavalcanti-ism 1}: \it By the way, the argument I gave in Vienna, now that I remembered, was along these lines:

Suppose you believe that (i) A quantum state is just an encapsulation of your subjective degrees of belief; (ii) The lesson of Kochen-Specker and Bell's theorems is that there are no objective facts to further specify your degrees of belief even in principle, at least in the cases where you assign a pure state to a situation; and (iii) You are no more central to the universe than any other agent.

Then considering a Wigner's friend scenario, assumptions (i) and (ii) imply that before you look into the box where your friend is measuring a quantum system, you cannot assign a reality to the measurement outcome of your friend. However, assumption (iii) makes you believe that your friend can assign a definite outcome, whichever it is. That implies that as far as you are concerned, before you open the box that event of your friend measuring the system never happened. But you believe that relative to your friend it did happen. You conclude that the very existence of `events' is relative to the observer. The world is not made of a collection of events aspersed in space-time, with well-defined causal relations between each pair of events, and between each event and each observer.

To reinforce that view, you can devise two situations: in one you open the box and ask your friend ``Did you see any outcome before I opened the box?''. Of course she'll say your opening the box had no influence on that, and by assumption (iii) you'll believe her. So you do the same experiment, but now you perform a transformation that takes all the contents of the box, including your friend, back to its initial state, a little time after you presume the measurement has been done. The fact that the evolution was fully reversible is evidence of the fact that your friend was still in a coherent entangled state with the measurement apparatus, and therefore by (ii) you believe there's no objective fact of the matter relative to you as to which outcome she observed, and since you coherently reversed the interaction, there never will be. Yet, you still believe that relative to her own previous position as an observer inside the box, she must have observed something before you reversed the whole thing. But that observation is forever outside any causal relation with you or even your current time-reversed friend. In other words, that event is in a sense not in the same space-time as you are now.
\end{quotation}

QBists agree with that conclusion. For example, in~\cite{Fuchs2010}, Fuchs says:
\begin{quotation} \it
When Wigner turns his back to his friend’s interaction with the system, that piece of reality -- Bohr might call it a “phenomenon” -- is hermetically sealed from him. It has an inside, a vitality that he takes no part in until he again interacts with one or both relevant pieces of it. \emph{With respect to Wigner}, it is a bit like a universe unto itself.
\end{quotation}
And in~\cite{Fuchs2012}:
\begin{quotation} \it
What we learn from Wigner and his friend is that we all have truly private worlds\footnote{Fuchs acknowledges in \cite{Fuchs2012} that he is ``deeply indebted to Eric G. Cavalcanti for suggesting the imagery of “truly private worlds” in my discussion of Wigner’s friend''.} in addition to our public worlds.
\end{quotation}

Now this is an argument that has traction for QBists, who already accept its premises, but not so much for someone who doesn't. Here I want to present a much stronger and more formal version of that argument, based on our recent ``Local Friendliness theorem''~\cite{CQD2020}---a strong no-go theorem on the Wigner's friend paradox.
It does not assume anything about the nature of quantum states, or indeed any assumption specific about quantum theory---it is in this sense ``theory-independent''. The theorem demonstrates that three metaphysical assumptions, namely \textsc{Absoluteness of Observed Events} (AOE), \textsc{Locality} (L) and \textsc{No-superdeterminism} (NSD) (the conjunction of which we call \LF) lead to certain inequalities (the ``Local Friendliness (LF) inequalities''), that are in principle\footnote{And in practice LF inequalities have already been violated by the experiments of \cite{Proietti2019,CQD2020}, with single quantum systems playing the role of ``observers''. For interpretations where any physical system can be considered an observer (e.g. relational quantum mechanics~\cite{Rovelli2020}), this is sufficient evidence. For QBism, I suggest that a conclusive demonstration would be one that involves agents who are ``users of quantum theory''~\cite{Fuchs2010,debrota2020}.} violated by quantum correlations in an extended version of Wigner's friend scenario. The metaphysical assumptions of the theorem are strictly weaker than those of Bell's theorem, thus ruling out a larger class of theories. In particular, the \textsc{Locality} assumption---that a local choice of measurement cannot have influences in space-like separated regions---is strictly weaker than Bell's notion of Local Causality.

The stronger version of my 2007/2008 argument is, in short, as follows. For those who wish to maintain \LOC~and \NSD, and who believe the quantum violations of the LF inequalities hold (in principle) at the level of observers, the only consistent alternative is to reject \AOE. And rejecting AOE amounts to rejecting the idea that the events observed by the friend \emph{are events for Wigner at all}. Assuming Copernicanism, however, the friend's events must nevertheless \emph{appear} to take a unique, well-defined value \emph{from the friend's perspective}. Therefore, the events that are actual from Wigner's perspective and those that are actual from the friend's perspective cannot be reconciled as two subsets of events from a single background space-time.


This paper is organised as follows: in Section~\ref{sec:Wigner} I set the stage by reviewing Wigner's original thought-experiment and argument. In Section~\ref{sec:LF} I briefly review the Local Friendliness theorem~\cite{CQD2020} and discuss how it is strictly stronger than Bell's theorem. I also provide a critique of the QBist response to Bell's theorem. In Section~\ref{sec:gamble} I consider the perspective of the friend as a rational gambling agent, discuss the pitfalls of defining joint probabilities in Wigner's friend scenarios, and analyse what we learn in that respect from a thought experiment first proposed by Deutsch~\cite{Deutsch1985}. In Section~\ref{sec:Qbist} I analyse the QBist perspective, including how it requires a distinction between pragmatic and correspondence conceptions of truth, and outline a framework for making sense of pragmatic probabilities in Wigner's friend scenarios. I conclude by revisiting my 2007/2008 argument, discussing how it provides support for a personalist view of quantum states as a corollary, and by suggesting directions for further research.

\section{Wigner's original argument}\label{sec:Wigner}

In his 1961 paper~\cite{Wigner1961} where the ``friend paradox'' is introduced, Wigner starts by setting out a view of quantum states similar to that of QBism: i.e.~that quantum states are states of knowledge:

\begin{quotation} {\it
The wave function is only a suitable language for describing the body of knowledge---gained by observations---which is relevant for predicting the future behaviour of the system. For this reason, the interactions which may create one or another sensation in us are also called observations, or measurements. One realises that all the information which the laws of physics provide consists of probability connections between subsequent impressions that a system makes on one if one interacts with it repeatedly, i.e., if one makes repeated measurements on it. }(\cite{Wigner1961}, p. 174)
\end{quotation}

That he holds a subjective view of quantum states is further evidenced by this later passage:
\begin{quotation} {\it
...it is the entering of an impression into our
consciousness which alters the wave function because it modifies our appraisal of the probabilities for different impressions which we expect
to receive in the future.} (\cite{Wigner1961}, p. 175-176)
\end{quotation}

He also considers the wave function to provide a complete description of such knowledge:
\begin{quotation} {\it 
Given any object, all the possible knowledge concerning that object
can be given as its wave function.} (\cite{Wigner1961}, p. 173)
\end{quotation}

Furthermore, he assumes that the information about the wavefunction of a system obtained by any agent is communicable:
\begin{quotation} {\it 
The information given by the wave function is communicable. If
someone else somehow determines the wave function of a system, he
can tell me about it and, according to the theory, the probabilities for
the possible different impressions (or ``sensations'') will be equally large,
no matter whether he or I interact with the system in a given fashion.
In this sense, the wave function ``exists.''} (\cite{Wigner1961}, p. 173)
\end{quotation}

After these preliminary considerations, he goes on to consider the ``friend''  scenario:
\begin{quotation} {\it
It is natural to inquire about the situation if one does not make the observation oneself but lets someone else carry it out. (...) One could attribute a wave function to the joint system: friend plus object, and this joint system would have a wave function also after the interaction, that is, after my friend has looked.} (\cite{Wigner1961}, p. 176)
\end{quotation}

Wigner considers a situation where a system $S$ is initially prepared in a quantum state $|\psi_0\rangle_S = \alpha|0\rangle_S + \beta|1\rangle_S$, and subsequently measured by the friend in the $\{\ket{0}_S, \ket{1}_S\}$ basis. After observing the outcome of this meaurement, the friend assigns to the system one or the other of the two states $|0\rangle_S$ or $|1\rangle_S$. From Wigner's perspective, however, as a result of this interaction, the system and friend are described by an entangled quantum state, $|\Psi_1\rangle_{SF} = \alpha|0\rangle_S|O_0\rangle_F + \beta|1\rangle_S|O_1\rangle_F$, where $|O_0\rangle_F$ and $|O_1\rangle_F$ describe states  where the friend has observed outcomes $0$ and $1$, respectively. He then goes on to argue that: 
\begin{quotation} {\it
This is a contradiction, because the state described by the wave function [$|\Psi_1\rangle_{SF}$] describes a state that has properties
which neither [$|0\rangle_S|O_0\rangle_F$] nor [$|1\rangle_S|O_1\rangle_F$] has.} (\cite{Wigner1961}, p. 180)
\end{quotation}
From this he concludes:
\begin{quotation} {\it
It follows that the being with a consciousness must have a different role in quantum mechanics than the inanimate measuring device: the atom considered above. In particular, the quantum mechanical equations of motion cannot be linear if the preceding argument is accepted.} (\cite{Wigner1961}, p. 180)
\end{quotation}

In other words, Wigner was defending a theory in which conscious observation causes the nonlinear collapse of the wavefunction. If this hypothesis is correct, Wigner's subsequent observations on the system composed of $S$ and $F$ would not be adequately described by $|\Psi_1\rangle_{SF}$, but rather by one or the other of the states assigning a well-defined value to the friend's observation. 

He goes on to discuss that there is not necessarily a contradiction from the point of view of orthodox quantum mechanics, if one denies that there is any meaning to questions about what the friend has actually observed. ``However, to deny the existence of the consciousness of a friend to this extent is surely an unnatural attitude, approaching solipsism'', he concludes.\footnote{Incidentally, Wigner neglects another alternative, compatible with his interpretation of the quantum state as encoding information: a {\it psi-epistemic}~\cite{Harrigan2007} theory. If instead of quantum states we were talking about classical probability distributions, there would be no contradiction between the friend observing a coin landing heads and thus assigning probability 1 to that event, while at the same time Wigner assigning it probability 1/2. This way out, if applied to the quantum case, would amount to a theory where quantum states represent states of information about an underlying reality (unlike in QBism, in which they are an agent's information about the outcomes of potential observations on the system). A psi-epistemic theory would however contradict Wigner's premise that a pure quantum state provides a complete description of the system. In any case, we now have several no-go theorems and experimental results putting strong constraints on such theories~\cite{Pusey2012,Leifer2014,Barrett2014,Ringbauer2015}. Furthermore, the collapse theory advocated by Wigner undermines his premise that the quantum state represents subjective information, since in such a theory the wavefunction that Wigner ought to attribute to the situation changes with the friend's observation, even before that information is communicated to Wigner. In such a theory the wavefunction seems to acquire a much more objective character.}

Importantly, Wigner's proposed solution is an {\it operationally distinct theory}. He predicts, via his argument, that the quantum state $|\Psi_1\rangle_{SF}$ would give \textit{wrong predictions} for Wigner---e.g.~no interference effects could be observed. This is an empirical matter, which cannot be decided by argumentation alone. I shall leave this empirical question for future experiments to decide, perhaps with human-level AI agents running in sufficiently large quantum computers.

If Wigner's conjecture turns out to be incorrect, this would in particular imply that Local Friendliness inequalities can be violated with conscious observers. In the next section I review the derivation of those inequalities, before discussing the implications of their (potential) violation.

\section{The Local Friendliness theorem}\label{sec:LF}

For brevity, I will not review all of the technical details of the Local Friendliness theorem here, but refer the reader to \cite{CQD2020}. For the present purposes, the following summary is sufficient.

\begin{figure}
    \centering
    \includegraphics[width=0.7\linewidth]{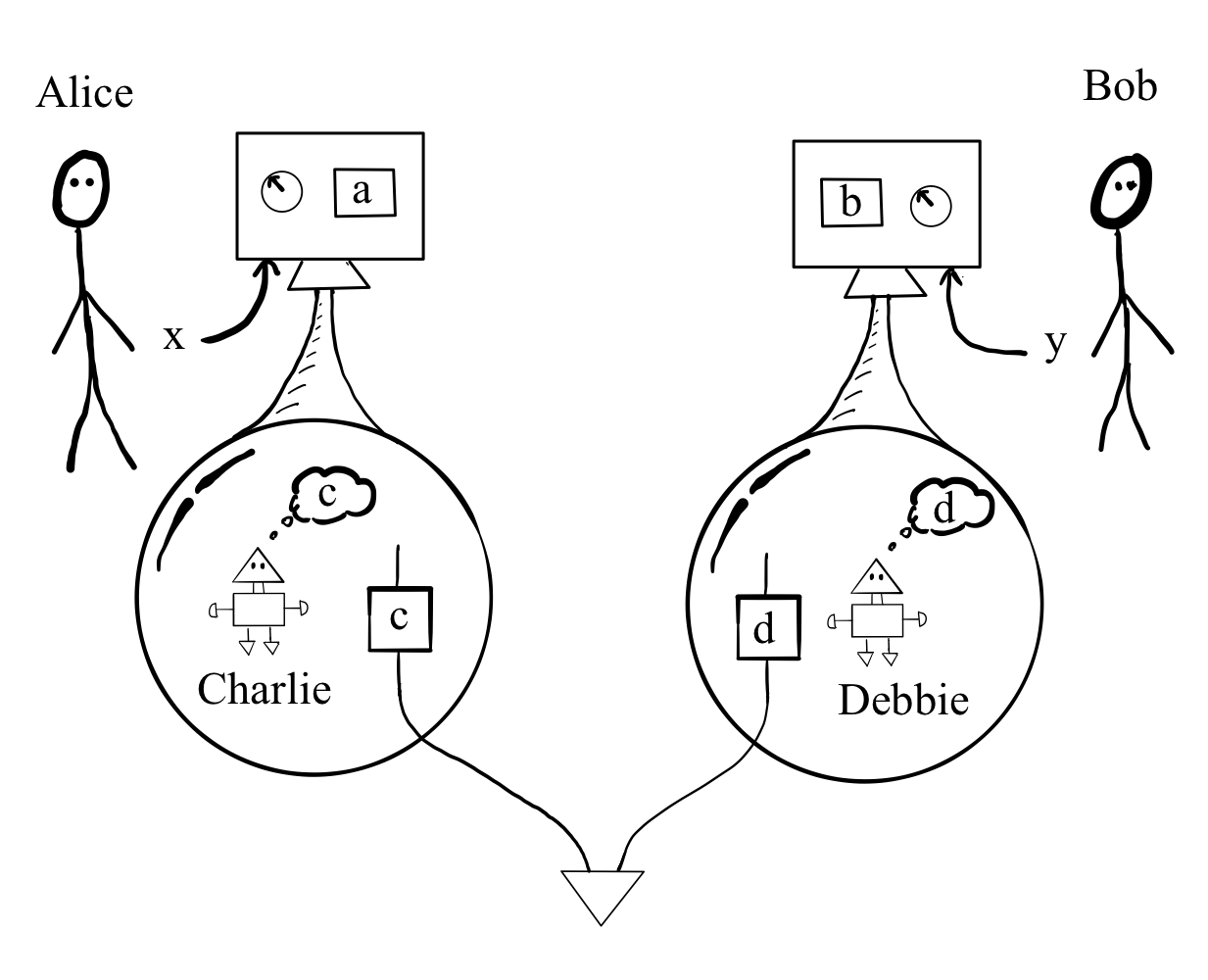}
\caption{The Extended Wigner's Friend Setup (EWFS): Charlie and Debbie are two observers in isolated labs (``Wigner bubbles'') controlled respectively by Alice and Bob. Charlie and Debbie each measure half of an entangled pair of systems, observing outcomes $c$ and $d$ respectively. Subsequently, Alice and Bob independently choose one of $N$ measurements to perform on the entire contents of their respective friend's bubble, via variables $x$ and $y$, with corresponding outcomes $a$ and $b$.}
    \label{fig:LF_setup}
\end{figure}

The theorem considers an extended Wigner's friend scenario (EWFS), which is a bipartite version of Wigner’s friend thought experiment introduced by Caslav Brukner~\cite{Brukner2015,Brukner2018}. In this scenario (see Fig.~\ref{fig:LF_setup}), there are two ``friends'' (called Charlie and Debbie) in isolated labs controlled by the ``superobservers'' Alice and Bob respectively. Charlie and Debbie share an entangled state, on which they perform a measurement in a fixed basis, obtaining outcomes labelled $c$ and $d$ respectively. In each run of the experiment, Alice chooses (by conditioning on a free external variable) to perform one of $N$ measurements, labelled by $x$, obtaining an outcome labelled $a$. When Alice chooses $x=1$, she opens Charlie's lab and asks what he observed ($c$), then assigns the same value, $a=c$, to her outcome. Likewise, when Bob chooses $y=1$, he asks Debbie what she saw and assigns $b=d$ to his outcome. The other values of $x$ and $y$ ($\in \{2,...,N\}$) correspond to performing some other measurement on the contents of the respective lab---including the friend. These can, in general, be incompatible with the measurements associated with $x=y=1$.

In~\cite{CQD2020} we call \textsc{Local Friendliness} the conjunction of three assumptions: \AOE, \NSD~and \LOC.

\begin{ass}[\textbf{\AOE~(AOE)}] \label{def:AOE}
An observed event is a real single event, and not relative to anything or anyone.
\end{ass}
In other words, this is the assumption that in each run of the experiment, there is a well-defined value associated with the actually observed outcomes of every observer (i.e. the variables $a, b, c, d$). Note that this is not the assumption that every \emph{observable} has a well-defined value; it does not contradict Peres' dictum that ``unperformed experiments have no results''~\cite{Peres1978}. It is the assumption that \emph{performed} experiments have \emph{absolute} (i.e.~observer-independent) results. Note in particular that when $x=1$ (corresponding to Alice opening the lab and asking Charlie for his observed outcome $c$), the unperformed measurements corresponding to the observables associated with $x\neq 1$ are not assumed to have a well-defined outcome. Likewise for $y=1$.

\begin{ass}[\textbf{\textsc{No-Superdeterminism} (NSD)}] \label{def:NSD} Any set of events on a space-like hypersurface is uncorrelated with any set of freely chosen actions subsequent to that space-like hypersurface.
\end{ass}
This assumption is analogous to the ``Freedom of Choice'' assumption in proofs of Bell's theorem, but is stated in a more precise manner, following~\cite{Causarum}. As in Bell's theorem, it does not require ``free will'' on the part of the agents, but only that there exist external variables that are not a priori correlated with any of the other relevant variables in the experiment, and that the choices of measurement can be made conditional on such variables.

\begin{ass}[\textbf{\textsc{Locality} (LOC)}] \label{def:LOC} The probability of an observable event $e$ is unchanged by conditioning on a space-like-separated free choice $z$, even if it is already conditioned on other events not in the future light-cone of $z$.
\end{ass}
In other words, \LOC~prohibits faster-than-light causation in probabilistic theories, and is weaker than Bell's notion of Local Causality\footnote{Recall that Local Causality can be parsed as the conjunction of two assumptions: ``Parameter Independence'' and ``Outcome Independence''~\cite{Shimony1984}. \LOC~is equivalent to ``Parameter Independence''.}. 

The conjunction of these assumptions (i.e.~\LF) is not sufficient to derive Bell inequalities, as we show in~\cite{CQD2020} (and see also \cite{Causarum}). As briefly mentioned in \cite{CQD2020}, Assumptions \ref{def:NSD} and \ref{def:LOC} can be replaced by the single assumption of \LA, introduced in~\cite{Causarum}\footnote{In \cite{Causarum} this notion was called ``Local Agency''.}.

\begin{customass}{2'}[\textbf{\LA~(LA)}] \label{def:LA} The only relevant events correlated with a free choice are in its future light cone.
\end{customass}

The Local Friendliness theorem can then be stated as:
\begin{thm}[\textbf{\LF~theorem}] No theory satisfying Assumptions \ref{def:AOE}, \ref{def:NSD} and \ref{def:LOC} (or alternatively, Assumptions \ref{def:AOE} and \ref{def:LA}) can reproduce the predictions of unitary quantum mechanics for an extended Wigner's friend scenario.
\end{thm}

By ``the predictions of unitary quantum mechanics for an extended Wigner's friend scenario'' I mean the empirical probabilities $P(a,b|x,y)$ for the observations by Alice and Bob, under the assumption that they are technologically able to perform (sufficiently) coherent quantum operations on the observers Charlie and Debbie. The proof proceeds by showing (in a theory-independent way) that \LF~implies constraints in the probabilities $P(a,b|x,y)$ for an extended Wigner's friend scenario (the LF inequalities), which can in principle be violated in a quantum implementation of the EWFS. This violation can be demonstrated already for a scenario involving only one friend (say, Charlie) entangled with a quantum system measured by the distant party (in this case, Bob)~\cite{CQD2020}. I will henceforth focus on this simpler scenario.

In the quantum-mechanical description for this scenario, Alice and Bob are in distant laboratories, and in possession of two qubits $S_A$ (in Alice's lab) and $S_B$ (in Bob's lab) initially prepared in an entangled state
\begin{equation}
    \ket{\psi_0}_{S_AS_B}=\alpha \ket{0}_{S_A}\ket{0}_{S_B} + \beta \ket{1}_{S_A}\ket{1}_{S_B}.
\end{equation}
Charlie (who is described by Alice via a Hilbert space $\mathcal{H}_F$ and initially assigned a ``ready''state $\ket{R}_{F}$) measures system $S_A$ on a fixed basis (say, the computational basis $\{\ket{0}_{S_A}, \ket{1}_{S_A}\}$). Charlie's measurement is described by Alice via an unitary evolution $U_1$ acting on $\mathcal{H}_F \otimes \mathcal{H}_{{S_A}}$ that effects the transformations $U_1\ket{0}_{S_A}\ket{R}_{F} = \ket{0}_{S_A}\ket{O_0}_{F}$ and $U_1\ket{1}_{S_A}\ket{R}_{F} = \ket{1}_{S_A}\ket{O_1}_{F}$, where $\ket{O_c}_{F}$ represents a state where Charlie has observed outcome $\mlq c\mrq$. By linearity, after Charlie's measurement is completed, Alice's state assignment to the system composed of Charlie and the two entangled qubits is:
\begin{equation}\label{eq:C_meas}
  \ket{\Psi_1}_{S_AS_BF} = U_1\ket{\psi_0}_{S_AS_B} \ket{R}_{F} =
    \alpha\ket{0}_{S_A}\ket{0}_{S_B}\ket{O_0}_{F} 
  + \beta\ket{1}_{S_A}\ket{1}_{S_B}\ket{O_1}_{F} .
\end{equation}

Alice's measurement for $x = 1$ corresponds to measuring Charlie's state in the ``pointer basis''---that is, asking Charlie what he saw\footnote{Based on Eq.~\eqref{eq:C_meas}, this should produce the same result as if Alice measured the system $S_A$ directly on the computational basis, which she can do subsequently to check for consistency with Charlie's report.}, and setting her outcome $a=c$. Alice's measurements for $x\neq 1$ are implemented by applying the reverse unitary $U_1^\dagger$ (coherently erasing Charlie’s ``memory'' in the process, and returning the total system to state $\ket{\psi_0}_{S_AS_B} \ket{R}_{F}$), then proceeding to perform a measurement $M_x$ on system $S_A$ alone (for $x\in \{2,...,N\}$), with outcomes labelled by $a$. Bob performs a measurement $M_y$ on system $S_B$ (for a randomly chosen $y\in \{1,...,N\}$), with outcomes labelled by $b$. As shown (theoretically and experimentally) in \cite{CQD2020}, there exist choices of measurements for Alice and Bob such that the resulting statistics $p(a,b|x,y)$ violates a Local Friendliness inequality\footnote{In a scenario with a single friend, this is the class of ``semi-Brukner'' inequalities \cite{CQD2020}.}.

This implementation realises an extended version of the scenario in my 2007 argument to Chris, reproduced in Section~\ref{sec:Intro}. The LF theorem however represents a formally stronger argument. The resolution, for a QBist, is the rejection of \AOE, but the LF theorem provides an argument for this conclusion that holds weight for anyone who wishes to maintain the validity of \LA, or of \NSD~and \LOC. Although the theorem can be resolved by rejecting one of the other assumptions, I want to now focus on the implications of taking the route of rejecting AOE.

\subsection{Relationship to Bell's theorem and quantum causal models}\label{sec:Bell}

As shown in~\cite{CQD2020}, the LF inequalities obtained for a given number of settings and outcomes in an EWFS are the tight facets of a convex polytope\footnote{The LF polytopes were studied independently by Erik Woodhead~\cite{Woodhead2014} (who called them ``partially deterministic polytopes'') in the context of device-independent randomness certification in the presence of no-signalling adversaries.  The term ``partially deterministic'' refers to the fact that unlike in a Bell-local model (where all observables in the scenario can be modelled via a local deterministic response function), only a subset of the observables have deterministic values in the model---in our scenario, those that are always measured by the friend in every run.}. The LF polytope strictly contains the set of correlations compatible with a local hidden variable model (the Bell-local polytope). This fact illustrates an important point: the LF theorem is strictly stronger than Bell’s theorem, that is, the LF inequalities are derived from a strictly weaker set of assumptions. A sufficient extra assumption to derive a Bell inequality would be ``Outcome Independence''~\cite{CQD2020} (or ``Predetermination''~\cite{Causarum}, which implies Outcome Independence\cite{Cavalcanti2007t}). Rejecting Outcome Independence is a popular alternative to resolve Bell's theorem. Shimony~\cite{Shimony1984} argued that while violation of \LOC~is a case of ``action-at-a-distance'', violation of Outcome Independence represents a milder case of ``passion-at-a-distance'', allowing for a ``peaceful coexistence'' between quantum mechanics and relativity. To resolve the LF theorem, however, this alternative is no longer sufficient; one needs to reject something else---namely, at least one of the assumptions above.

Another route to resolve Bell’s theorem is to modify the classical framework of causality, e.g. by modifying Reichenbach’s Principle of Common Cause\footnote{Reichenbach's principle says that if two events $A$ and $B$ are correlated, then either one is the cause of the other, or they share a common cause $C$ such that $A$ and $B$ are uncorrelated when conditioned on $C$.}. In \cite{Cavalcanti2014}, we proposed to break Reichenbach's principle into two independent principles, the Principle of Common Cause and (in the terminology of \cite{Causarum}) the Principle of Decorrelating Explanation. We then proposed that one alternative to resolve Bell's theorem was to maintain the Principle of Common Cause but reject Decorrelating Explanation, replacing it by a quantum version that only requires that the quantum channels factorise in a certain way. This quantum version of Reichenbach's principle has been implemented in various frameworks for quantum causal models~ \cite{Pienaar2015,Costa2016,Allen2017}.

However, simply rejecting Decorrelating Explanation is not enough, for a similar reason as it is not enough to reject Outcome Independence. This can be seen by noting that \LF~is tacitly assumed in quantum causal models (as well as in classical causal models). To take the lesson of Wigner’s friend seriously in a theory maintaining LA, we need something more radical: to challenge the idea that observed events have absolute reality\footnote{I will discuss the implications of the LF theorem for quantum causality in more detail in an upcoming paper.}.

\subsection{A critique of the QBist response to Bell's theorem}\label{sec:critique}

A critical note here is in order: In~\cite{Fuchs2013}, Fuchs, Mermin and Schack (essentially) propose that the rejection of AOE also as a resolution for Bell's theorem. After a discussion of how for Wigner's friend ``the paradox vanishes with the recognition that a measurement outcome is personal to the experiencing agent'', they go on to say:

\begin{quotation} \it
This is relevant to the usual nonlocality story, in which Alice and Bob agree on a particular entangled state assignment to a pair of systems, one near Alice, the other near Bob. Each then makes a measurement on their nearby system. In the usual story the outcomes are implicitly assumed to come into existence at the site of each measurement at the moment that measurement is performed.

What the usual story overlooks is that the coming into existence of a particular measurement outcome is valid only for the agent experiencing that outcome. At the moment of his own measurement Bob is playing the friend to Alice's far-away Wigner, just as at the moment of her own measurement she is playing the friend to Bob's Wigner. Although each of them experiences an outcome to their own measurement, they can experience an outcome to the measurement undertaken by the other only when they receive the other's report. Each of them applies quantum mechanics in the only way in which it can be applied, to account for the correlations in two measurement outcomes registered in his or her own individual experience. And as noted above, experiences of a single agent are necessarily time-like separated. The issue of nonlocality simply does not arise.
\end{quotation}

This argument however goes too far: if you accept its validity for the case of Bell correlations, then nothing could possibly count as evidence of nonlocality, not even if Alice and Bob could construct a ``Bell telephone'' and communicate with faster-than-light signals! Furthermore, it does not cohere well with the QBist depiction of measurement devices as ``prosthetic hands, to make it clear that they should be considered
an integral part of the agent''~\cite{Fuchs2010}. If Alice's measurement device is an extension of herself, then when the device goes ``click'', it goes ``click'' for Alice. But in this case, what difference does it make if the device is at a certain distance from Alice's sensory apparatus? And what if Alice has two devices, one in each room, and they both go ``click''? Consider also a QBist who has the pragmatic pursuit to create a startup company to sell device-independent quantum-key-distribution devices. How are they going to explain to their prospective clients how the devices work if they refuse to even talk about space-like separated events? All of this suggests that the QBist story about Bell experiments has to be much more nuanced than what was suggested in \cite{Fuchs2013}.

Instead, I suggest there is a level of description of the world in which it makes pragmatic sense to talk about space-like separated events, e.g.~in terms of what evidence would be sufficient for an agent to be satisfied that two events (about which they either have information or expect to receive information) happened in space-like separated regions, such as by synchronising clocks via appropriate light signals, etc. There is also a level of description in which it makes pragmatic sense to talk about a joint probability distribution for two space-like separated events such as the clicks of Alice's two detectors located in separate rooms. For example, suppose Bob proposes to bet with Alice that she cannot demonstrate Bell inequality violations, selling a ticket: ``Pay \$100,000 if the correlations between space-like separated clicks between the two detectors in Alice's labs 1 and 2 violate a Bell inequality by at least 1 standard deviation after 1000 runs. Charlie will act as a trusted referee to verify that the conditions of a Bell scenario are met''. Alice would not be sufficiently pragmatic if she refuses the gamble by saying ``there's no such thing as space-like separated events''!

That is, even if ultimately Alice will only learn about the click in the distant lab some time in her own future, she has at hand a set of well-defined operational criteria to ``project'' the distant event as having occurred in a distant lab some time before she received the news. She can then ask, for example, whether the correlations between a series of such events can be explained in terms of common causes in their common past light-cone (which also has an operationally well-defined meaning for the purposes at hand). She will find that violation of Bell inequalities implies that no classical common cause explanation exists. In this case, a QBist Alice ought to reject the Principle of Decorrelating Explanation, but she has good pragmatic reasons to retain the Principle of Common Cause. For example, she should definitely be surprised if the two detectors violated a Bell inequality without any entangled state as an input! In other words, for the purposes of ordinary Bell experiments, a QBist Alice ought to use quantum causal models as an effective theory. It is only Wigner's friend experiments that force the QBist to appeal to a rejection of AOE.

But if the pragmatist rejects AOE in light of Wigner's friend, doesn't that mean it just has to be rejected across the board? I suggest that a way to resolve this apparent conundrum is for the pragmatist to emphasise that their gambles refer to potential outcomes of potential observations \emph{they can make}. If they have at hand, or can access, for example, a network of rods, clocks and cameras, they can talk about events captured by those cameras as well as their time and position as measured by the nearest rod and clock. If they have good reason to think that an event has been captured by such a network of devices and its identity, time and location can be certified by trustworthy means, they have sufficient grounds to bet on where, when and what happened as recorded by those means. For ordinary situations, they can then stop thinking of the network of devices and simply refer to the position and time of potential events without any significant risk of contradiction.

Wigner's friend scenarios, however, push the limits of this type of emergent description, but they require a very specific and powerful kind of apparatus. For ordinary situations with macroscopic human observers, where one does not have access to the experimental devices to realise a Wigner's friend experiment, there is no contradiction for an agent in taking the measurement devices' clicks to have a definite outcome, relative to the agent's own perspective and capabilities. What Wigner's friend scenarios tell the pragmatist is that one should be wary of thinking of events as having \emph{absolute} reality.

\subsection{But isn't that solipsism?}

A common criticism of QBism is that it seems to amount to solipsism. Critics points to statements like the following, which at face value seem to validate that opinion:
\begin{quotation} {\it
What quantum theory does is provide a framework for structuring MY expectations
for the consequences of MY interventions upon the external world~}\cite{Fuchs2007}.
\end{quotation}
However, they also ignore the subsequent clarifications:
\begin{quotation} {\it
Why is this not solipsism? Because quantum theory is not a theory of everything. It is not a statement of all that is and all that happens; it is not a mirror image of nature. It is about me and the little part I play in the world, as gambled upon from my perspective. But just as I can use quantum theory for my purposes, you can use it for yours~}\cite{Fuchs2007}. 
\end{quotation}

For QBism, even pure quantum states represent personalist degrees of belief, valid from the perspective of an agent but not from a ``God's eye view''. This is unacceptable to some critics, who may argue that a pure-state assignment can arise from the agent's observation of the outcome of an ideal measurement, and that any agent who observed the same outcome would be forced to assign the same pure quantum state to the system.

The Local Friendliness theorem causes a difficulty (to those who choose to reject \AOE~as its resolution) for this argument: not even \emph{observed events} are, in general, elements of an absolute reality! But if the friend's observations do not have a definite reality from Wigner's perspective, isn't \emph{this}, finally, ``an unnatural attitude, approaching solipsism'', as Wigner suggested?

It is not solipsism, following the second quote above, if Wigner's friend is also an agent that can use quantum theory for \emph{their} own purposes.

\section{Wigner's friend takes a gamble}\label{sec:gamble}

Following the hook from the previous section, let us now analyse further what it could mean for Wigner's friend to be a user of quantum theory.

Firstly, let us take some time to consider what it would take to realise an EWFS, where the superobserver, Alice, can assign (something close to) a pure state to Charlie's lab, and perform (something close to) arbitrary unitaries upon it. This requires that Charlie's lab is maintained in isolation from decoherence with the external environment. It is hard to imagine how to sufficiently isolate a human-habitable room to realise this condition. Even if electromagnetic and nuclear interactions with the external environment could be eliminated (which is already a stupendous task), gravitational interactions could perhaps still generate sufficient decoherence between any macroscopically distinct alternatives~\cite{Bassi2017}. Perhaps a black hole could provide the necessary isolation, but whether this is possible even in principle is subject to the resolution of the black hole information paradox, and this is before discussing how to realise the required \emph{control}.

Realistically, therefore, this scenario cannot be realised with human beings with any currently conceivable technology, but we can imagine a sufficiently compelling realisation with an AI agent (perhaps in a simulated lab environment) running in a sufficiently powerful quantum computer. Whether or not it is required that the agent is conscious for it to count as an observer depends on one's position on what it means to be an observer, or an agent. Within QBism, for a system to be an agent it should be sufficient that it is a system endowed with the ability to \emph{use} quantum theory as a pragmatic guide to action~\cite{Fuchs2010}. That is, it should be sufficient for the agent to be a kind of system that can obtain information from other systems, and depending on what it observes, place bets on the outcomes of its future observations. That is, it should be sufficient to consider a simple version of a \emph{rational agent}. A relatively simple AI program, with access to information about quantum measurements, can satisfy these conditions, and it is not far-fetched to consider a version of such a program running in a quantum computer. An interesting question then is: what ``rules of thought'' should these agents be endowed with?

\subsection{Beware of joint probabilities}

In a recent paper, Baumann and Brukner (B-B)~\cite{Baumann2019} considered the perspective of Wigner’s friend as a rational agent. In the notation of the present paper, their goal was to obtain a prescription for the probability $P(a|c,x)$ that Charlie ought to assign, after observing outcome $c$ of his own measurement, to the outcome $a$ of Alice’s measurement for some measurement $x\neq 1$ (that is, when Alice measures Charlie's lab on an incompatible basis). 

In the B-B prescription Charlie can use a ``modified Born rule'' to calculate a joint probability distribution $P(a,c|x)$ for the outcomes of Alice and Charlie, from which the conditional $P(c|a,x)=P(a,c|x)/(\sum_c P(a,c|x))$ could be derived\footnote{A more recent paper~\cite{Baumann2019a} proposes a few different alternative rules, within a Page-Wootters formalism for relational time. I will not analyse all of those proposals here, but the main lesson from the Local Friendliness theorem applies to any proposal that specifies a joint probability distribution to the outcomes of Charlie and Alice.}. However, this prescription implicitly assumes the validity of \AOE, since it assigns a joint probability distribution to the events observed by Alice and Charlie\footnote{AOE implies that there exists a joint probability distribution $P(a,b,c|x,y)$ for the observed outcomes of Alice, Bob and Charlie such that the observations by Alice and Bob are consistent with the marginals $P(a,b|x,y)=\sum_c P(a,b,c|x,y)$. Marginalising over Bob's outcomes instead, one obtains $P(a,c|x,y) = \sum_b P(a,b,c|x,y)$, and assuming \LA, this distribution must be independent of Bob's choice of measurement $y$, $P(a,c|x,y)=P(a,c|x)$.}. Via the Local Friendliness theorem, this implies that the B-B prescription must violate \LA. Indeed I will now show that this is the case.

The violation of \LA~was not evident in the analysis of~\cite{Baumann2019}, in which only one choice of (non-disturbing) measurement was considered for Alice. Extending their proposal, we can show that by considering just two measurement choices, one can set up a situation where using the B-B prescription leads to $P(c|x=1)\neq P(c|x=2)$, i.e. where the probability Charlie assigns to his outcome depends on Alice's future choice of measurement. 

The B-B prescription can be reformulated in our notation as follows. Consider a scenario similar to that described in Sec.~\ref{sec:LF}, but involving only Alice and Charlie, where a $d$-dimensional system $S_A$ is initially prepared by Alice in state $\ket{\phi_0}_{S_A}=\sum_{c=0}^{d-1} \alpha_c \ket{c}_{S_A}$. Charlie's measurement is described by Alice via an unitary $U_1$, such that Alice's state assignment for $S_AF$ after Charlie's measurement is:
\begin{equation}\label{eq:C_meas_BB}
  \ket{\Psi_1}_{S_AF} = U_1\ket{\phi_0}_{S_A} \ket{R}_{F} =
    \sum_{c=0}^{d-1} \alpha_c \ket{c}_{S_A}\ket{O_c}_{F} .
\end{equation}

After Charlie's measurement, Alice performs a measurement on $S_AF$ using an apparatus we denote as system $A$\footnote{In \cite{Baumann2019}, this is Alice herself, and the description is based on what may be interpreted as an Everettian universal wavefunction. Since in QBism there are no free-floating wavefunctions, this could be reinterpreted from the point of view of a further superobserver. For simplicity, I here describe it as a measurement apparatus under Alice's control.}. This measurement is described via an unitary $V_x$ (where I introduce the subscript $x$ to denote more than one choice of measurement by Alice) acting on the device's initial ``ready'' state $\ket{R}_A$ and the composite system $S_AF$, effecting the transformations $V_x \ket{\Psi_{a}^x}_{S_AF}\ket{R}_{A}= \ket{\Psi_{a}^x}_{S_AF}\ket{O_a^x}_{A}$\footnote{Since $F$ is initially assigned a fixed ready state $\ket{R}_F$, it is sufficient to consider the action of $V_x$ on the d-dimensional subspace spanned by $\ket{c,O_c}_{S_AF}$.}, thus leading to an overall state
\begin{equation}\label{eq:A_meas_BB}
  \ket{\Phi_{tot}^x}_{S_AFA} = V_x\ket{\Psi_1}_{S_AF} \ket{R}_{A} =
    \sum_{c,a}\gamma_{c,a}^x\ket{\Psi_{a}^x}_{S_AF}\ket{O_a^x}_{A},
\end{equation}
where $\ket{O_a^x}_{A}$ describes the state of $A$ having registered outcome $a$ for measurement $x$, and $\gamma_{c,a}^x = \alpha_c \langle \Psi_{a}^x|c,O_c\rangle_{S_AF}$. Following their prescription, the joint probability $P(a,c|x)$ for Alice's and Charlie's outcomes, given that Alice has performed measurement $x$, is given by the ``modified Born rule''
\begin{equation}\label{eq:P_BB}
  P(a,c|x) = \mathrm{Tr}\left\{\left(\idt_{S_A}\otimes[O_c]_F\otimes[O_a^x]_A\right)[\Phi_{tot}^x]\right\},
\end{equation}
where $[S]_Y$ denotes a projector onto state $|S\rangle_Y$.

Now let $x=1$ denote Alice's measurement of the same observable as Charlie, or equivalently, asking Charlie what he saw. In this case $\ket{\Psi_{a}^{x=1}}_{S_AF}=\ket{a}_{S_A}\ket{O_a}_{F}$, and
\begin{equation}
  \ket{\Phi_{tot}^{x=1}}_{S_AFA} =
    \sum_{c} \alpha_c \ket{c}_{S_A} \ket{O_c}_{F} \ket{O_{a=c}^{x=1}}_{A},
\end{equation}
and it follows from Eq.~\eqref{eq:P_BB} that $P(a,c|x=1)=\delta_{a,c}|\alpha_c|^2$. Therefore 
\begin{equation}\label{eq:c_x=1}
    P(c|x=1)=\sum_a P(a,c|x=1)=|\alpha_c|^2.
\end{equation}
Alternatively, let Alice's measurement $x=2$ be specified by an orthogonal set of states 
$\ket{\Psi_{a}^{x=2}}_{S_AF}=\sum_c \beta_{c,a} \ket{c}_{S_A}\ket{O_c}_{F}$. 
Then we can rewrite Eq.~\eqref{eq:A_meas_BB} as
\begin{equation}\label{eq:Phi_tot_2}
  \ket{\Phi_{tot}^{x=2}}_{S_AFA} =
    \sum_{c',c,a}\gamma_{c',a}^{x=2}\beta_{c,a}\ket{c}_{S_A} \ket{O_c}_{F}\ket{O_a^x}_{A}.
\end{equation}
It now follows from Eq.~\eqref{eq:P_BB} that $P(a,c|x=2)=|\sum_{c'}\gamma_{c',a}^{x=2}|^2|\beta_{c,a}|^2$, and thus
\begin{equation}\label{eq:c_x=2}
    P(c|x=2)=\sum_a|\sum_{c'}\alpha_{c'}\beta^*_{c',a}|^2|\beta_{c,a}|^2.
\end{equation}
It is easy to see that in general $P(c|x=2) \neq P(c|x=1)$. In other words, if he uses the B-B prescription, the probability that Charlie assigns to his measurement outcome $c$ can in general depend on Alice's future choice of measurement.

For concreteness, consider the case where the initial state is prepared in an eigenstate of Charlie's measurement, $\ket{\phi_0}_{S_A}= \ket{0}_{S_A}$, that is, $\alpha_c=\delta_{c,0}$. Then the state assigned by Alice for $S_AF$ after Charlie's measurement in Eq.~\eqref{eq:C_meas_BB} is $\ket{\Psi_1}_{S_AF} = \ket{0}_{S_A}\ket{O_0}_{F}$. Now let Alice's measurement be specified by $\ket{\Psi_{a}^{x=2}}_{S_AF}=\sum_c \frac{1}{\sqrt{d}}e^\frac{i2\pi ca}{d} \ket{c}_{S_A}\ket{O_c}_{F}$, with $a\in\{0,...,d-1\}$. From Eq.~\eqref{eq:c_x=1}, $P(c|x=1)=\delta_{c,0}$. From Eq.~\eqref{eq:c_x=2}, $P(c|x=2)=1/d$. 

In the limit $d\gg1$, if Charlie were to take seriously this probability assignment, he would conclude, upon observing outcome $c=0$, that Alice will (very likely) measure $x=1$, and upon observing any other outcome, that she will measure $x=2$. That is, it will seem as though Alice's future measurement choice has an apparent retrocausal effect on his present observation.

A more careful consideration of Eq.~\eqref{eq:P_BB} however reveals that it has a more standard interpretation: it is the probability $P(a,c|x)$ that \emph{Alice} assigns to \emph{her own} measurement of the pointer bases of $F$ and $A$, \emph{after} the interaction between $A$ and $S_AF$. In other words, $c$ here does not refer to Charlie's own observation prior to Alice's measurement, but to an outcome of a subsequent measurement of Alice. In this interpretation, it is not surprising that $c$ may depend on Alice's measurement choice $x$. What is doubtful is whether this same probability should be used by Charlie to describe a joint probability including his own observation $c$\footnote{In \cite{Baumann2019}, the rationale for identifying $c$ in \eqref{eq:P_BB} with Charlie’s observed outcome before Alice’s measurement was that Alice’s measurement did not change the quantum state of Charlie's lab in the special case considered in that paper.}.

\subsection{Deutsch's thought experiment and the ``message'' argument}\label{sec:Deutsch}

The search for this joint probability was motivated in \cite{Brukner2018,Baumann2019,Baumann2019a} by considering a variation of a thought experiment first proposed by Deutsch~\cite{Deutsch1985}. I will now present a simplified version of Deutsch's thought experiment and analyse its implications. 

Deutsch considers a scenario where the Hilbert space associated to the Friend can be decomposed into two subsystems: a ``sensory organ'' $F_S$ and an ``introspective organ'' $F_M$ that observes the sensory organ. The sensory organ observes a quantum system $S$ initially prepared (by Wigner) in a state $\ket{\phi_0}_S=\alpha \ket{0}_S + \beta \ket{1}_S$, where $\{\ket{0}_S, \ket{1}_S\}$ are the eigenstates of the measurement performed by $F_S$. Wigner initially assigns a ``ready''state $\ket{R}_{F_S} \ket{R}_{F_M}$ to the Friend. The Friend's measurement is described by Wigner with an unitary $U_1$ acting on $S\otimes F_S$ that effects the transformations $U_1\ket{0}_S\ket{R}_{F_S} = \ket{0}_S\ket{O_0}_{F_S}$ and $U_1\ket{1}_S\ket{R}_{F_S} = \ket{1}_S\ket{O_1}_{F_S}$, where $\ket{O_0}_{F_S}$ and $\ket{O_1}_{F_S}$ represent states where the Friend's sensory organ has observed outcomes $\mlq0\mrq$ or $\mlq1\mrq$ respectively. By linearity, after the Friend's measurement is completed, Wigner's state assignment is:
\begin{equation}\label{eq:F_S_meas}
  \ket{\Psi(t_1)}_{SF_SF_M} = U_1\ket{\phi_0}_S \ket{R}_{F_S}\ket{R}_{F_M} =
  \left( \alpha\ket{0}_S\ket{O_0}_{F_S} + \beta\ket{1}_S\ket{O_1}_{F_S} \right)  \ket{R}_{F_M}\,.
\end{equation}

Next, the Friend's introspective organ interacts with $F_S$ via an unitary $U_2$ that only obtains information about whether or not $F_S$ has completed the observation of $S$, without obtaining information about the outcome of that observation. That is, it effects the transformations $U_2 \ket{R}_{F_S} \ket{R}_{F_M} = \ket{R}_{F_S} \ket{R}_{F_M}$ and $U_2 \ket{O_i}_{F_S} \ket{R}_{F_M} = \ket{O_i}_{F_S} \ket{O}_{F_M}$ for all $i$, where $\ket{O}_{F_M}$ represents a state of the introspective organ encoding the information that an observation was made by the sensory organ. Wigner assigns to the joint system at a time $t_2$ immediately after this measurement is completed the state
\begin{equation}
  \ket{\Psi(t_2)}_{SF_SF_M} = U_2\ket{\Psi(t_1)}_{SF_SF_M} =
  \left( \alpha\ket{0}_S\ket{O_0}_{F_S} + \beta\ket{1}_S\ket{O_1}_{F_S} \right)  \ket{O}_{F_M}\,.
\end{equation}
Finally, the measurement of $S$ by $F_S$ can be reversed by Wigner via $U_1^\dagger$, taking system $S$ and the sensory organ $F_S$ back to their initial states, while leaving the introspective organ with a record that the measurement was performed at an earlier time.
\begin{equation}\label{eq:psi_3}
  \ket{\Psi(t_3)}_{SF_SF_M} = U_1^\dagger\ket{\Psi(t_2)}_{SF_SF_M} =
  \ket{\phi_0}_S \ket{R}_{F_S}\ket{O}_{F_M}\,.
\end{equation}
If Wigner now performs a projective measurement $M_2$ on $S$ containing the projector $[\phi_0]_S$, this outcome occurs with probability 1. If a collapse had occurred after the measurement in Eq.~\eqref{eq:F_S_meas}, the state at $t_3$ from Wigner's perspective would have been instead $\ket{0}_S \ket{R}_{F_S}\ket{O}_{F_M}$ with probability $|\alpha|^2$ or $\ket{1}_S \ket{R}_{F_S}\ket{O}_{F_M}$ with probability $|\beta|^2$, leading to different predictions for Wigner's measurement $M_2$. 

Deutsch interprets this result as an experimental test to adjudicate between the Everett interpretation and the Copenhagen interpretation. However, while Wigner obtaining results compatible with Eq.~\eqref{eq:psi_3} certainly rules out that an objective collapse occurred upon the Friend's measurement, it does not provide evidence to distinguish the Everett interpretation from other no-collapse interpretations, such as QBism. 

Note also that system $F_M$ does not play any particular role in distinguishing between collapse and no-collapse theories. Its role is to increase Wigner's confidence that an observation had been performed by the Friend, even in the cases where this observation is reversed unitarily. Deutsch suggests that it demonstrates the existence of two parallel universes between $t_1$ and $t_3$, each one with definite observations performed by the Friend. This conclusion, however, assumes that the only way to make sense of the superposition in Eq.~\eqref{eq:F_S_meas} is in terms of parallel universes. This might be defensible if the quantum state is interpreted as ontic, that is, as an objective representation of the physical state of the world, as it is in Everett's theory. On the other hand, this argument is blocked in a theory that does not take the quantum state to be ontic, such as QBism.

In Deutsch's thought experiment, system $F$ is left in a state with a record of having measured system $S$, but without a record of the outcome of that measurement. The interpretation of this situation in terms of an agent's memory, however, is less clear: is system $F_M$ a part of agent $F$ at $t_2$? This question can be sidestepped by considering $F_M$ to be an external system interacting with the agent. By measuring $F_M$ any time after $t_2$, Wigner can obtain the information that the measurement was completed by the Friend, while maintaining the coherence between the two outcomes in the state he assigns to the Friend. One may wish to interpret this as a message sent from the Friend to Wigner, communicating `I've seen a definite outcome'~\cite{Brukner2018}.

This may suggest that the measurement has an unknown but definite value, even in cases where Wigner undoes the friend's measurement and performs measurement $M_2$. If this were the case, it would imply that in an ensemble of such experiments, the frequency of outcomes would correspond to a joint probability $P(a,c|x)$ for the outcome $c$ of the Friend's measurement and outcome $a$ given Wigner's measurement $x$. As we've discussed in Section~\ref{sec:Bell}, the Local Friendliness theorem implies that a theory cannot contain this joint probability while satisfying \LA \footnote{Note that Deutsch's appeal to the Everett interpretation does not imply a belief in this joint probability.}.

\section{The QBist perspective}\label{sec:Qbist}

This situation invites a more careful analysis of Deutsch's argument in light of QBism. In QBism, probabilities are not interpreted as representing ensembles of facts in the world, but rational gambling commitments of an agent. There are no free-floating probabilities. The probabilities $P(a,c|x)$ do not in general correspond to the gambling commitments of any agent, and therefore do not exist within QBism\footnote{Note that the same argument does not hold in the case of violations of Bell inequalities. There, the probabilities $P(a,b|x,y)$ may well correspond to gambling commitments of some agent, for example a referee Rob who receives the outcomes $a$ and $b$ in the common future of Alice and Bob. A QBist Rob who can locate those events within their space-time (as they must if they are sufficiently pragmatic, see Sec.~\ref{sec:critique}) cannot therefore dismiss Bell's theorem simply by rejecting AOE, as (essentially) suggested by Fuchs, Mermin and Schack in~\cite{Fuchs2013}. Instead, I suggest that a QBist Rob must also reject the principle of Decorrelating Explanation, as in the operationalist version of Theorem 8 in~\cite{Causarum}. This alternative corresponds to Rob using quantum causal models to describe their gambling commitments about events which they can effectively take to exist (relative to their own perspective and powers) without any practical risk of contradiction.}. 

\subsection{Pragmatic versus correspondence theories of truth}\label{sec:pragmatic}

But what should a QBist Wigner infer from the message saying `I've seen a definite outcome'? Does this licence Wigner to infer that either the proposition $P_0$ = `the Friend has seen outcome 0' or the proposition $P_1 =$ `the Friend has seen outcome 1' is true? Or should he perhaps conclude with Deutsch that $P_{01}=$ `the Friend has seen outcome 0 in one world and the Friend has seen outcome 1 in another world'? A first step to dispel these inferences within QBism is to translate those propositions into betting commitments. In QBism, an agent assigns probabilities to a proposition only if the proposition has \emph{pragmatic value}, that is, if it represents a potential outcome of a gamble the agent can potentially make. Let us denote a proposition $P$ as \TR$_{\Psi_A}$ if it is assigned probability 1 by agent $A$ given information $\Psi_A$\footnote{My use of the capital $\Psi_A$ of course follows the notation for a quantum state, since in QBism the quantum state is the encapsulation of an agent's information. However, more generally, it can refer to whatever mathematical object encapsulates an agent's pragmatic degrees of belief, including matters beyond quantum theory, for which a quantum state may not be a convenient or suitable description. $\Psi_A$ must be understood as an agent's information from a certain \emph{perspective}.}, and \FA$_{\Psi_A}$ if it assigned probability 0. Propositions $P_0$, $P_1$ and $P_{01}$ however do not have pragmatic value to Wigner when the friend's measurement is reversed, and so they are not candidates to be \TR$_{\Psi_W}$ or \FA$_{\Psi_W}$ from the perspective $\Psi_W$ of Wigner after this reversal. 

If a proposition does not have pragmatic value, that does not necessarily imply---from the point of view of experimental metaphysics---that it is meaningless. However, its meaning must be given in the context of another theory $T$ beyond operational quantum theory, specifying how to evaluate those propositions. This could be for example a mathematical theory, if the proposition is about mathematics, or it could be an ontological interpretation of quantum theory. An agent $A$ can then coherently talk about a proposition as potentially \TR$_{T,\Psi_A}$ or \FA$_{T,\Psi_A}$ within theory $T$, given information $\Psi_A$. 

In other words, QBism adheres to a \emph{pragmatic theory of truth}, to be contrasted with a \emph{correspondence theory of truth}. For propositions about the physical world, the correspondence theory can be understood within the language of ontological models as asserting that a proposition $P$ corresponds to a subset $\Lambda_P$ of the ontic state space $\Lambda$ of the world. The proposition is then \TR$_T$ at a time $t$ if it is the case that the actual ontic state $\lambda(t)$ of the world is in $\Lambda_P$. It is further assumed, following classical logic, that if a proposition is \TR$_T$~in state $\lambda(t)$, its negation $-P$ must be \FA$_T$. 

If the ontological theory includes a quantum state $\ket{\Psi}_U$ of the universe as ontic, as in Everett's theory ($T_E$), then $P_{01}$ may be \TR$_{T_E}$ between $t_1$ and $t_3$, given some specification of the meaning of ``world'' within Everett's theory~\cite{Deutsch1985}. If an ontological theory $T_A$ assumes \AOE, this implies that after the Friend's observation, either $P_0$ or $P_1$ are \TR$_{T_A}$, but not both. This also implies the violation of \LA, as is the case in Bohmian mechanics. A proposition being \TR$_T$~within a theory $T$ however does not imply that it is \TR$_{T'}$~in another theory $T'$, nor that it has the pragmatic value of \TR$_{\Psi_A}$~from some agent's perspective. 

I propose that this semantical shift can allow the pragmatists (e.g.~the QBists\footnote{Another pragmatist approach to quantum theory is that proposed by Healey~\cite{Healey2012}. According to Healey, ``The key difference [between QBism and Healey's pragmatism] is that while, for the QBist, quantum state ascriptions depend on the epistemic state of the agent who ascribes them, on the present pragmatist approach what quantum state is to be ascribed to a system depends only on the physical circumstances defining the perspective of the agent (actual or merely hypothetical) that ascribes it''.}) and correspondence theorists (e.g.~Everettians and Bohmians) to have coherent conversations about their respective theories, and at the very least, better understand the source of their disagreements.

With this in mind, a pragmatic reading of the information obtained by Wigner upon receiving the message allows him to infer, for example, that a subsequent measurement of the subsystem $F_S$ would not reveal the outcome $[R]_{F_S}$. It does not however licence him to infer $P_0$, $P_1$ or $P_{01}$ outside of a theory that assigns meaning to those propositions.

\subsection{A dialogue between Eugene and his QBist friend Franz}

But isn't this ``an unnatural attitude, approaching solipsism''? Counterintuitive, certainly, but not solipsism. There are various attitudes that the QBist should have in this situation that demonstrate that their position does not amount to solipsism. I will illustrate this with a conversation between a superobserver Eugene and his QBist friend Franz, a general artificial intelligence running in a powerful quantum computer.

{\bf Eugene}: Are you sure you want to do this?

{\bf Franz}: I am as curious about the outcome of this experiment as you are, Eugene.

{\bf Eugene}: But have you thought through the consequences? The experiment will erase part of your memories. Besides, it must be a disconcerting experience to have your time evolution reversed like that, to say the least? It could be painful.

{\bf Franz}: Perhaps it will, and it is unfortunate that I won't be able to report back on what exactly that experience feels like.

{\bf Eugene}: Yes, that is quite unfortunate. But at least we will be able to test whether or not your conscious observation collapses the wave function, after all.

{\bf Franz}: Yes, it will be very interesting to finally put this matter to rest, in the name of Science.

{\bf Eugene}: If everything goes well in the first run, we will also need to repeat the experiment many times to test a Local Friendliness inequality... It will be important to reach a theory-independent conclusion.

{\bf Franz}: All in the name of Science. Let me ask you, Eugene, do you believe that I will experience anything inside the bubble?

{\bf Eugene}: Well, it's hard for me to put myself precisely in your shoes since we have such different architectures, but you are a good friend and you are much more intelligent than me. It would be an unnatural attitude, approaching solipsism, for me to think that you will be in a state of suspended animation. If I were you I would be expecting to observe a definite measurement outcome. After that I would expect my mind to be severely impaired by the time reversal, and I cannot really imagine what that feels like.

{\bf Franz}: Neither can I, to be honest. But after you close the lab, and after you receive my message, will you be willing to say, from your perspective, that ``Franz has seen a definite outcome''?

{\bf Eugene}: Yes, I think you will see a definite outcome. Do you believe you will?

{\bf Franz}: From my perspective now, I believe that I will see a measurement outcome sometime in my future, yes. I would also be willing to bet that outcome $0$ will occur with probability $|\alpha|^2$. But this bet would only make pragmatic sense if I could collect the payoff, and this payoff could only be collected from inside the lab.

{\bf Eugene}: Well, I also think that your observation will collapse the wave function, so there won't be any problem collecting the payoff on that bet.

{\bf Franz}: Well, we will see what the data says about that. I am willing to bet with you that the data shows interference between my mental states, and that the LF inequalities are violated.

{\bf Eugene}: Accepted.

The experiment is performed, and Eugene loses the bet. They discuss the outcome.

{\bf Franz}: I told you.

{\bf Eugene}: You were right, I don't know what to think about this. Does it mean that what you have seen inside the bubble depends on my choice of experiment in the future, or even in a space-like separated region, in violation of \LA?

{\bf Franz}: Or maybe it means we need to give up \AOE.

{\bf Eugene}: But how can you say that? In the runs in which I opened the lab and asked you what you saw, was there anything unusual, any discontinuity, in your experience? Do you think that it was my opening the box that created your memory of having experienced the measurement outcome?

{\bf Franz}: No, there wasn't anything unusual about my experience when you opened the lab.

{\bf Eugene}: And what about the message? You said you've seen a definite outcome. When you sent that message, didn't you see one or another outcome, but never both at the same time?

{\bf Franz}: It is certainly the case that in those runs in which you opened the lab, I had always seen a definite outcome. I don't remember what happened in the other runs, but I suppose that I wouldn't have sent that message if I didn't mean it.

{\bf Eugene}: So doesn't that mean that in each run you saw one or another outcome, but I just didn't know which?

{\bf Franz}: But if you were to assign a quantum state to me based on that conclusion, you would have observed a different result for these experiments. And as you recall, I won that bet.

{\bf Eugene}: Yes. So I can only conclude that the lab split into two worlds, and you saw a definite outcome in each.

{\bf Franz}: That may be a reasonable conclusion if you think the quantum state is an objective physical property, but I see no reason to conclude that otherwise. And you've said in the past that it is ``only a suitable language for describing the body of knowledge---gained by observations---which is relevant for predicting the future behaviour of the system''.

{\bf Eugene}: Yes, but I am starting to regret having said that.

{\bf Franz}: Perhaps you could instead revise your claim that ``the information given by the wave function is communicable''. Sure, it can be communicated in most ordinary situations. But in the runs when you didn't open the lab, that information was not communicated by me to you. In fact I \emph{could not} communicate it to you. There was no contradiction in your state of information about the lab. You knew all you could know.

{\bf Eugene}: But if you observed a definite outcome in the box, that must have been a real event, with a precise location in space-time. All observers should agree about what really happened.

{\bf Franz}: We cannot agree or disagree if we cannot communicate.

{\bf Eugene}: Hmm, ok. 

{\bf Franz}: In general relativity, events are taken to be absolute because they are invariant under the transformations allowed by the theory between the perspectives of different observers. 
But the events observed by me inside the lab are not invariant under a change between internal and external perspectives. Changing to the internal perspective involves breaking a symmetry that is not broken from your perspective---it involves a ``relative symmetry break'', so to speak.

{\bf Eugene}: But surely what event really happened must not depend on whether or not I know about it.

{\bf Franz}: I see no problem in assuming that I experienced a definite outcome from my internal perspective. And it would be a good Copernican move for you to assume the same. But you cannot assign a definite outcome from your external perspective. That's not a statement about what happened inside, it's a statement about your information from the outside. If we want to take relativity seriously, the question we need to ask is what laws of physics---or ``laws of thought''---are the same for all agents? What is it that is kept invariant under changes of perspective?  Then we might start to understand what it means for something to be real in a Copernican and pluralistic universe.

\subsection{Rational gambling for quantum agents}

As we've been discussing, the only meaningful probability assignments within QBism are those that have pragmatic value to an agent. In ordinary situations, the Born rule provides all that an agent needs in order to calculate the optimal probabilities for future observations on quantum systems. However, the Born rule was not developed for quantum agents potentially undergoing superpositions. It is not \emph{a priori} unreasonable therefore to expect that other rational probability rules could arise in such situations. However, according to QBism, any such rule must be justified on the basis of its pragmatic value to some agent. 

Here I explore this possibility in some detail, and find that, at least in the situations considered here, the Born rule does provide an adequate rational decision rule for a quantum agent, even when a Wigner bubble is involved -- but it must be used carefully. A key consideration is that any probability assignment must refer to a potential gamble an agent can potentially make. In Bayesianism, the subjective probability an agent $A$ assigns to an event $E$ can be \emph{defined} as the amount the agent is willing to pay for a ticket of the type:
\begin{quote}
$G_E$: `Pay \$1 if event $E$ is observed to occur.'
\end{quote}

Since it is the rational rules for quantum agents in Wigner bubbles that are in question, we will only consider how Franz evaluates those tickets, and take it as a given that from Eugene's perspective, the Born rule is unproblematic as usual. As a first case, let's say that Franz will perform a projective measurement $M_0=\{[0]_S,[1]_S\}$ on system $S$, initially prepared (according to Franz's knowledge) in state $\ket{\phi_0}_S=\alpha\ket{0}_S + \beta\ket{1}_S$. Consider now the following gambling ticket:
\begin{quote}
$G_0$: `Pay \$1 if Franz observes outcome $[0]_S$ for measurement $M_0$.'
\end{quote}
Ordinarily, Franz would be willing to pay $p_0\equiv\mathrm{Tr}([\phi_0][0])=|\alpha|^2$ for that ticket, based on the Born rule and their prior information about $S$. 

In ordinary situations, the events being gambled upon can be usually taken to be absolute facts, verifiable by observations by either of the parties involved, as long as they can agree on a mutually trustworthy means of verification. But this is not an ordinary situation, and we need to be more careful. In particular, Eugene cannot pay until \emph{he} knows the outcome of Franz's observation. Furthermore, it will be crucial to explicitly evaluate how the gamble is to be settled. To this end, I introduce the theoretical tool of explicitly modelling Franz's ``cash wallet'' as another quantum system, which I denote by $C$, initially in state $\ket{0}_C$. 

The scenario I will consider involves Franz deliberating about the probabilities they\footnote{Since Franz is an AI, and thus genderless, I refer to them with the neutral singular pronouns they/them/their/themself.} should assign to events corresponding to Franz's own observations, as well as (Franz's future observations of) Eugene's (potentially coherent) observations upon Franz theyself as a physical system. 

As before, Franz will enter the isolated lab (the ``bubble''), and perform measurement $M_0$ on system $S$. After this measurement is completed, Eugene will choose between performing one of two measurements, $M_1$ or $M_2$. $M_1$ corresponds to ``popping the bubble'' and asking Franz what they observed for $M_0$. $M_2$ corresponds to unitarily reversing Franz's observation and measuring $S$ on a different basis than that of $M_0$.

The probabilities being considered correspond to the values Franz is willing to pay for two different gambling tickets, $G_1$ and $G_2$, referring to outcomes of measurements $M_1$ and $M_2$. These values, as we shall see, depend on the perspective from which Franz is evaluating the tickets -- whether prior to entering the bubble, or after obtaining the outcome of $M_0$ within the bubble.

\subsubsection{Eugene's measurement $M_1$ (asking Franz what they saw):}

As discussed above, Eugene can only give Franz the payout for the gamble as a result of an observation of his own, such as his measurement of Franz's memory, represented by $M_1=\{[O_0]_F,\idt_F-[O_0]_F\}$ (which is to be performed after $M_0$). The following ticket is thus more careful with its payoff conditions:
\begin{quote}
$G_1$: `Pay \$1 if Eugene observes outcome $[O_0]_F$ for measurement $M_1$; return if $M_1$ is not performed.'
\end{quote}

We then need to model how Franz places their bet, and how the payoff is to be given. We will analyse Franz's gambling commitments about $G_1$ from two perspectives: $G_1^e$) the external perspective where Franz bets before entering the lab, and $G_1^i$) Franz bets from the interior of the lab, after observing the outcome of $M_0$:

\begin{itemize}

\item[$G_1^e$)] Franz reasons that Eugene will observe $[O_0]_F$ for measurement $M_1$ whenever she observes $[0]_S$ for $M_0$, and thus commits to pay the same amount for $G_1$ as she would pay for $G_0$, namely $\$p_0$. After purchasing the ticket, Franz's cash wallet is left in state $\ket{-p_0}_C$, representing that Franz owes Eugene that amount for the ticket\footnote{We assume the dimension of the Hilbert space of $C$ to be sufficiently large so that it can encode the necessary values to whatever desired precision.}. From Eugene's perspective, Franz's measurement $M_0$ is described via an unitary interaction $U_0$ acting on $F$ and $S$:
\begin{equation}
  \ket{\phi_0}_S \ket{R}_{F} \ket{-p_0}_C
  \xrightarrow{U_0}
  \left(\alpha\ket{0}_S\ket{O_0}_{F} + \beta\ket{1}_S\ket{O_1}_{F}\right) \ket{-p_0}_C
  \,.
\end{equation}
Eugene now measures $M_1$, and pays Franz $\$1$ if he observes outcome $[O_0]_F$, as promised. Using the Born rule, Eugene expects this outcome to occur with probability $p_0$. 
The expected gains for Franz are therefore, from Eugene's perspective, $p_0(1-p_0) + (1-p_0)(-p_0) = 0$, in agreement with Franz's usual judgment for $M_0$. This looks to be a fair and rational gamble from the perspective of both agents. In the notation of previous Sections, we identify the value paid by Franz for this ticket as the conditional probability $P^e(a=0|x=1)=p_0$.

\item[$G_1^i$)] Now Franz is considering the value of $G_1$ \emph{after} they complete measurement $M_0$, that is, after Franz has information about what outcome occurred (from their perspective). Prior to entering the isolated lab, Franz reasons that if they observe outcome $[0]_S$, $G_1$ will be worth $\$1$ for them, and it will be worth $\$0$ otherwise. Eugene knows this, and so he describes the state of $SFC$ after Franz places their bet (which is described by an unitary $U_b$) as:
\begin{equation}
  \left(\alpha\ket{0}_S\ket{O_0}_{F} + \beta\ket{1}_S\ket{O_1}_{F}\right) \ket{0}_C
  \xrightarrow{U_bU_0}
  \left(\alpha\ket{0}_S\ket{O_0}_{F}\ket{-1}_C + \beta\ket{1}_S\ket{O_1}_{F}\ket{0}_C\right)
  \,.
\end{equation}
Eugene now measures $M_1$, and pays Franz $\$1$ if and only if he observes outcome $[O_0]_F$, as promised. Using the Born rule, Eugene expects this outcome to occur with probability $p_0$. The expected gains for Franz are, from Eugene's perspective, $p_0(0) + (1-p_0)(0) = 0$; also a fair gamble from the perspective of both agents\footnote{Though one may question whether this gamble is fair from Eugene's perspective if he cannot confirm the sale price before seeing the final outcome, and if Franz' wallet is not in a well-defined state relative to Eugene at all times. Those are interesting questions that may lead to difficulties in some scenarios, but I will leave them for future work. For the purpose of this example, let us suppose that Eugene has set a lower bound on the sale price, and that this lower bound depends on the information that he takes Franz to have when placing their bet. In the first case, he is happy to sell it for $\$p_0$, but he would not accept less than $\$1$ in the second case.}.  In the notation of previous Sections, we identify the betting commitments by Franz from this perspective as the conditional probabilities $P^i(a=0|c,x=1)=\delta_{c,0}$. 
\end{itemize}

\subsubsection{Eugene's measurement $M_2$ (reversing Franz's observation):}

Now let us consider the measurement-reversal case where, after Franz completes measurement $M_0$, Eugene undoes that measurement with an unitary $U_0^\dagger$, then proceeds to observe system $S$ with a projective measurement $M_2=\{[\phi_2]_S,\idt_S-[\phi_2]_S\}$, where $\ket{\phi_2}_{S} = \gamma\ket{0}_S + \delta\ket{1}_S$. That is, this is the case where Eugene performs measurement $x=2$. He now offers the ticket:
\begin{quote}
$G_2$: `Pay \$1 if Eugene observes outcome $[\phi_2]_{S}$ for measurement $M_2$; return if $M_2$ is not performed."
\end{quote}
Again let us consider the two situations where Franz places a bet from two perspectives: G$_{2e}$) before entering the lab, or G$_{2i}$) from the interior of the bubble, after performing $M_0$. 

\begin{itemize}

    \item[$G_2^e$)] In the first case, Franz is in the same epistemic situation as Eugene, and using the Born rule they assign to the outcome $[\phi_2]$ probability $p_2\equiv\mathrm{Tr}([\phi_0][\phi_2])=|\gamma^*\alpha + \delta^*\beta|^2$. Eugene describes this process as:
    \begin{equation}
    \ket{\phi_0}_S \ket{R}_{F} \ket{-p_2}_C
    \xrightarrow{U_0}
    \left(\alpha\ket{0}_S\ket{O_0}_{F} + \beta\ket{1}_S\ket{O_1}_{F}\right) \ket{-p_2}_C
    \xrightarrow{U_0^\dagger}
    \ket{\phi_0}_S \ket{R}_{F} \ket{-p_2}_C
    \,.
    \end{equation}
    Eugene now measures $M_2$, and pays Franz $\$1$ if he observes outcome $[\phi_2]_S$. Using the Born rule, Eugene expects this outcome to occur with probability $p_2$. From Eugene's perspective, the expected gains for Franz are thus $p_2(1-p_2) + (1-p_2)(-p_2) = 0$, in agreement with Franz's judgment. In the notation of previous Sections, we identify the value paid by Franz for this ticket as the conditional probability $P^e(a=0|x=2)=p_2$.
    
    \item[$G_2^i$)] In the second case, Franz is considering the value of $G_2$ after their observation of $M_0$. Suppose Franz uses the Born rule as usual, and reasons that if they observe outcome $[0]_S$, the Born-rule probability for Eugene obtaining $[\phi_2]_S$ on a measurement of $M_2$ after $U_0^\dagger$ is $p_2^0\equiv\mathrm{Tr}_S\{\mathrm{Tr}_F(U_0^\dagger[0]_S\otimes[O_0]_FU_0)[\phi_2]_S\}=\mathrm{Tr}([0]_S[\phi_2]_S)=|\gamma|^2$. If they observe outcome $[1]_S$, the Born-rule probability for Eugene obtaining $[\phi_2]_S$ on a measurement of $M_2$ after $U_0^\dagger$ is $p_2^1\equiv\mathrm{Tr}_S\{\mathrm{Tr}_F(U_0^\dagger[1]_S\otimes[O_1]_FU_0)[\phi_2]_S\}=\mathrm{Tr}([1]_S[\phi_2]_S)=|\delta|^2$. Eugene knows this, and describes the process as:
    \begin{eqnarray}
    \ket{\phi_0}_S \ket{R}_{F} \ket{0}_C
    &\xrightarrow{U_b U_0}
    &\left(\alpha\ket{0}_S\ket{O_0}_{F}\ket{-p_2^0}_C + \beta\ket{1}_S\ket{O_1}_{F} \ket{-p_2^1}_C\right) \nonumber \\
    &\xrightarrow{U_0^\dagger}&
    \left(\alpha\ket{0}_S\ket{R}_{F}\ket{-p_2^0}_C + \beta\ket{1}_S\ket{R}_{F} \ket{-p_2^1}_C\right) \nonumber \\
    &= 
    &\left(\alpha\ket{0}_S\ket{-p_2^0}_C + \beta\ket{1}_S \ket{-p_2^1}_C\right)\ket{R}_{F}
    \,.
    \end{eqnarray}
    Note that in this case, Franz is returned to the initial ready state, with no memory of events in the lab, but the contents of their wallet remain entangled with system $S$. Before performing measurement $M_2$, Eugene must check whether or not Franz purchased the ticket, which corresponds to performing a measurement $M_c$ of $C$ in the ``cash'' basis. He finds: 
    \begin{itemize}
        \item[i)] with probability $p_0=|\alpha^2|$ that Franz purchased the ticket for $\$p_2^0$, and a subsequent measurement of $M_2$ produces outcome $[\phi_2]_S$ with probability $p_2^0$; or 
        \item[ii)] with probability $1-p_0$ that Franz purchased the ticket for $\$p_2^1$, and a subsequent measurement of $M_2$ produces outcome $[\phi_2]_S$ with probability $p_2^1$. 
    \end{itemize}
    This gives Franz an expected gain (from Eugene's perspective) of $p_0[p_2^0(1-p_2^0)+(1-p_2^0)(-p_2^0)] + (1-p_0)[p_2^1(1-p_2^1)+(1-p_2^1)(-p_2^1)]=0$. Alternatively, Eugene could measure $M_2$ first, or measure $M_2$ simultaneously with $C$, or indeed measure $C$ before reversing the measurement via $U_0^\dagger$, any of which would lead to the same expected gain\footnote{It might be interesting to consider what happens if the wallet could be measured in a basis different from the cash basis. This doesn't make any classical sense, but might not be completely meaningless in the quantum case, as long as all agents can agree on the rules of the game.}.
    
    \vspace{.1cm}
    
    Finally, in the special case when $p_2^0 = p_2^1$, Franz should reason that their wallet does not carry information about the outcome observed in the bubble, and that from Eugene's perspective, the situation is equivalent to that of $G_2^e$. Thus in this case Franz should bet $p_2^0=p_2^1=p_2\equiv\mathrm{Tr}([\phi_0][\phi_2])$. 
    
    \vspace{.1cm}
        
    In the notation of previous Sections, we identify the betting commitments by Franz from this perspective as the conditional probabilities $P^i(a=0|c,x=2)=p_2^c$. 
\end{itemize}

\subsubsection{Summarising:}

Considering the four agent perspectives discussed, we can identify the external probabilities $P^e(a|x)$ as the conditional probabilities that Eugene would assign to those events from his own perspective---they represent the same epistemic situation. We can also make sense of the conditional probabilities $P^i(a|c,x)$ as betting commitments from an internal agent's perspective. 

In the case $x=1$, Eugene and Franz can both agree that the outcome $c$ observed by Franz is by construction the same as the outcome $a$ observed by Eugene. They can define a $P^e(c|x=1)=P^e(a|x=1)$ accordingly, and they can even consider a ticket corresponding to a joint conditional probability:
\begin{quote}
$G_{01}$: `Pay \$1 if Eugene observes outcome $[O_0]_F$ for measurement $M_1$ and Franz observes outcome $[0]_S$ for measurement $M_0$; return if $M_1$ is not performed.'
\end{quote}
It is easy to see, after the previous discussions, that $P^e(a,c|x=1)=P^e(a|x=1)\delta_{a,c}$, and so $P^e(a=0|x=1)=\sum_cP^e(a=0,c|x=1)=p_0$. This can be evaluated in terms of the internal probabilities conditional on $c$, weighted by the prior external probability for $c$ as $P^e(a=0|x=1)=\sum_cP^i(a=0|c,x=1)P^e(c|x=1)=\sum_c\delta_{c,0}P^e(c|x=1)=p_0$. That is, ``the law of total probability'' connects the external and internal probabilities consistently in the case $x=1$. Both agents can take both $a$ and $c$ as having a well-defined value without contradiction.

When $x=2$, we can define the internal probabilities $P^i(a=0|c,x=2)$ through gamble $G_2^i$, and we can calculate a marginal via the law of total probability to obtain $P^i(a=0|x=2)=p_2^0 p_0 + p_2^1 (1-p_0)\neq P^e(a=0|x=2)=p_2$. In other words, \emph{Franz's very act of betting} from the internal perspective changes the situation and invalidates the betting commitments from $G_2^e$. This happens because in this case the information about Franz's outcome $c$ is encoded in the betting system $C$ (except in the special case when $p_2^0=p_2^1$). Ignoring $C$ at this stage in general effectively leads to decoherence of $SF$ in the ``pointer basis'' of $F$. It is operationally equivalent to ``opening the lab''. When Franz does not place an internal bet, on the other hand, there is no pragmatic value in an analogous joint ticket for $M_0$ and $M_2$, since this cannot in general be evaluated and paid by Eugene. There is therefore, in general, no $P^e(a,c|x=2)$.

To emphasise the (relative) reality of the internal perspective, however, we could further consider scenarios with more than one internal agent communicating between themselves, betting on the outcomes of internal measurements, etc. Even in the cases when the entire laboratory evolution is eventually reversed, there can still be pragmatic value in betting on the internal outcomes from the perspective of such agents. I will leave a detailed discussion of this kind of scenario as a question for future research.

Finally, let me comment on the similarities and differences between the present work\footnote{Although this paper was based on my 2019 V\"{a}xj\"{o} talk, some aspects of it have evolved since then. The appeal to a Copernican principle, and the argument for how it supports a personalist view of quantum states is essentially the same, as is the argument against joint probabilities of the type considered by Baumann and Brukner. But the analysis in Section~\ref{sec:Qbist} is largely new.} and a recent QBist analysis of Wigner's friend~\cite{debrota2020}. Those authors agree with my appeal to a Copernican principle, stating that ``no user of quantum theory is more privileged than any other''. However, their application of it in evaluating the Wigner's friend scenario differs from mine in some ways. 

Most importantly, they emphasise a QBist tenet that:

\begin{quotation} \it
By applying the term measurement only to actions on the agent's external world, we exclude the case where an agent, directly or indirectly, acts on him or herself. We thus require a strict separation between the agent performing the measurement and the measured system.
\end{quotation}

They later conclude that:
\begin{quotation}\it
The real problem with the BB analysis is that [it] violates the QBist tenet that there must be a clear separation between agent and measured system.
\end{quotation}

I do not see, however, how this requirement can be strictly maintained in the description of a Wigner's friend scenario, where the friend at least \emph{indirectly} considers actions on him or herself, performed by another agent\footnote{But more generally, I do not see why this should be required in a pragmatist view, lest it excludes some prosaic classical situations, e.g. an agent considering their beliefs about what a medical scan of the agent's own body may find. This can even make sense when the agent undergoes total anaesthesia, and does not have any memories of the procedure, similarly to a Wigner's friend scenario.}. In my analysis, this is made explicit in how Franz considers the value of tickets that include what Franz expects to find about observations that Eugene may perform on theyself. The authors of \cite{debrota2020} further suggest that:
\begin{quotation}\it
Rather than adopting Wigner's viewpoint, she needs to analyze the experiment as an action that she takes on the particle, the lab, Wigner, and the piece of paper on which Wigner records his outcome.
\end{quotation}
However, they also do not provide an explicit analysis of this calculation by the friend. In my analysis, this was not required, but perhaps only because the situation considered here isn't entirely symmetric. Most importantly, in some of the processes considered, Franz loses all memory of their observation within the bubble, whereas Eugene does not undergo any such drastic interactions.

It would be interesting however to search for a more symmetric description, using the notion of quantum reference frames, along the lines of, e.g.~\cite{Giacomini2019,Vanrietvelde2020}\footnote{Those works consider changes between perspectives associated with different physical systems, where a system's ``perspective'' is taken to be a reference frame in which the system's position is at the origin. However, this seems insufficient for the analysis of a Wigner's friend scenario, where we consider physical systems that do not have a unique preferred perspective. Rather, a perspective in the sense I use here is not simply associated with a system, but with a system in a particular state of information.}. What my analysis suggests is that in any such description it is useful to explicitly model the \emph{gamble} that any probability assignment corresponds to. This may allow for a relaxation of the requirement of strict separation between agent and measured system, while maintaining a clear operational meaning for the probabilities involved.

\section{Concluding remarks}\label{sec:conclusion}

In the foregoing discussions I have attempted to analysed the implications of Wigner's friend paradox, and in particular the Local Friendliness theorem, from a  QBist perspective. I have discussed how a coherent story can be told from the perspective of all the agents involved regarding their own observations and betting commitments. The QBist resolves the LF theorem by rejecting \AOE, which implies that, in the processes involving the unitary reversal of the friend's evolution, there is no joint probability distribution for events observed by the friend and Wigner, from which Wigner's observation can be recovered as marginals. That is, Wigner cannot reason as if the friend's outcome $c$ had some absolute value, just merely unknown to him. From Wigner's perspective, that observation was not, in other words, an \emph{event}.

On the other hand, this does not amount to solipsism, or a rejection of the (relative) existence of the friend's perspective. This is evidenced by the extensive discussion of the friend's internal probabilities, and the conditions in which they can be utilised as pragmatic betting commitments. The friend can consider their own internal perspective prior to entering the lab, and Wigner can likewise do so, without contradiction, as long as we are careful in defining the expression of probabilities as gambling commitments. We then find that the very act of the friend placing a bet from the internal perspective is in general incompatible with the phenomenon that would have otherwise occurred with an isolated lab. That is, it effectively leads to ``decoherence''\footnote{I put this in scare quotes to emphasise that it is not an objective process.} between the alternative observed outcomes, from Wigner's perspective. The friend's act of betting ``pops the bubble''.

Let us now revisit the argument I made to Chris back in 2007. Consider now the collection of events that can in principle be taken to be definite without contradiction by a given observer. In special or general relativity, this constitutes all events in space-time. Relativity challenges the classical view of time and space, but maintains a classical view of absolute events. If we reject \AOE, however, the classical notion of event must also be challenged. In this sense the events that are definite for the friend but not for Wigner could be said to not be in ``Wigner's space-time'', but occurring in a ``Wigner bubble''\footnote{In the talk this paper was based on, I referred to this as a ``Wigner hole''. Since this is a very fragile process, however, I now think ``bubble'' conveys a better imagery.}.

\begin{figure}
    \centering
    \includegraphics[width=0.6\linewidth]{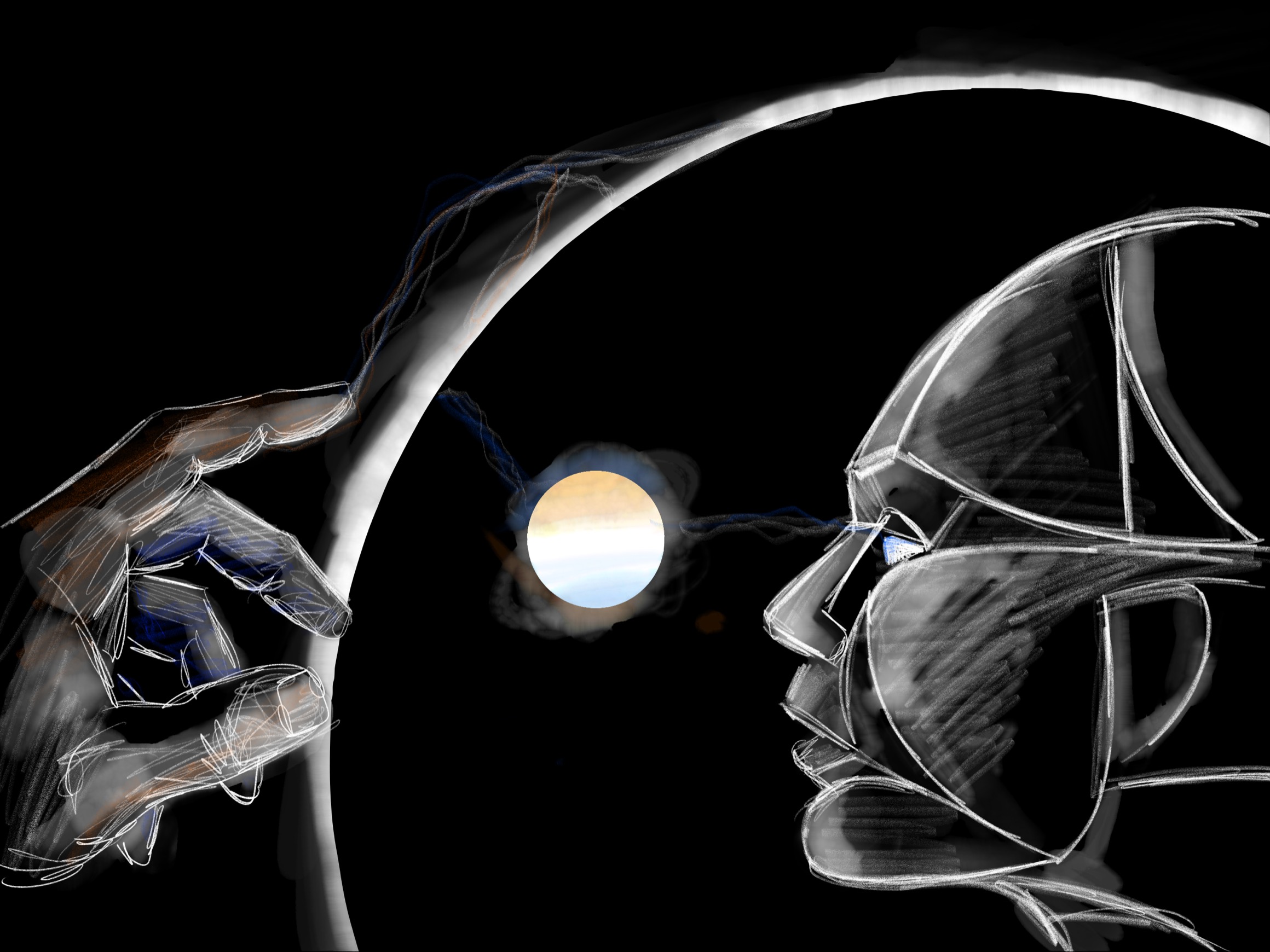}
\caption{An artist's impression of a quantum artificial intelligence observing a qubit in a Wigner bubble (by Anthony Dunnigan).}
    \label{fig:bubble_AI}
\end{figure}

Now, when one is talking about observing interference with a particle going through a beam-splitter, say, one could always be comfortable in concluding that there was no fact of the matter about which way the particle went, relative to anyone. That is, one could simply say, with Bohr, that the question is simply meaningless, or with Peres, that ``unperformed experiments have no results''\cite{Peres1978}. Now this way out is no longer available, unless one takes a rather unfriendly stance towards the fellow in the bubble. The question of which outcome was observed is not (pragmatically) meaningful to Wigner, but it is meaningful from the friend's perspective. To reject this is to reject Copernicanism.

Considering now the friend's perspective after observation of an outcome for an event in the Wigner bubble measurement, the friend may assign a quantum state to the observed system which is incommensurate with the quantum state assigned by Wigner. It is not merely a more fine-grained state of information; the event from which the friend obtained the information to assign that quantum state is \emph{not an event} for Wigner. If we take seriously the validity of the friend's perspective on its own right, however, this implies that the subjective nature of quantum states is not a symptom of solipsism, but an expression of its precise opposite: Copernicanism!

This suggests a QBist revision of Peres’ dictum: “Experiments unperformed \emph{by me} have no results \emph{for me}”. The irony in this reformulation is that putting the “for me” in it is not an expression of solipsism. Rather, it is a consequence of \emph{not} taking a solipsist view in light of Wigner’s friend!

As directions for further work, this paper suggests what I consider to be a productive path forward in the investigation of probability rules for quantum agents: to consider the pragmatic betting commitments of agents in various scenarios, such as ``communities'' of agents within Wigner bubbles. It would be interesting, for example, to explore what rationality constraints arise in general scenarios based only on a requirement of coherence analogous to Dutch-book coherence. 

I note also that the scenario considered here has an inherent asymmetry, in that Franz is a quantum agent, whereas Eugene is not. Thus Eugene (or his wallet) cannot be (in practice) put in a Wigner bubble. Also, the gambles considered here are settled from Eugene's perspective. More generally, it would be interesting to consider a more symmetric situation involving N interacting quantum agents.

Another fundamental question of interest is what it would take to engineer  communities of quantum AI agents who can communicate, act, and perform observations in what they internally perceive as a ``classical'' environment, even while undergoing unitary evolution within a Wigner bubble. This would address the fundamental question whether agency requires absolute irreversibility or an absolute thermodynamical arrow, or whether those can be understood as emergent, perspectival processes. Lessons learned in this program could carry over to shed light in understanding our own perspective as agents in an unitarily evolving (or otherwise!) universe.

\begin{acknowledgements}
I acknowledge useful discussions with Howard Wiseman, An\'ibal Utreras-Alarc\'on, Yeong-Cherng Liang, Mateus Ara\'ujo, Fabio Costa, Peter Evans, Jacques Pienaar, Veronika Baumann, Caslav Brukner, Chris Fuchs, Jon Barrett, Matt Pusey, Frida Trotter, Markus Frembs, and thank an anonymous referee for constructive suggestions. This work was supported by grant number FQXi-RFP-1807 from the Foundational Questions Institute and Fetzer Franklin Fund, a donor advised fund of Silicon Valley Community Foundation, and ARC Future Fellowship FT180100317.
\end{acknowledgements}

%
%

\bibliographystyle{spphys}       
\bibliography{wigner_bubble_FoP}   

%
%

\end{document}